\documentclass{aa}                           \usepackage{natbib,twoopt}
\usepackage[breaklinks=true]{hyperref}
\makeatletter
\newcommandtwoopt{\citeads}[3][][]{\href{http://adsabs.harvard.edu/abs/#3}%
  {\def\hyper@linkstart##1##2{}%
    \let\hyper@linkend\@empty\citealp[#1][#2]{#3}}}
\newcommandtwoopt{\citepads}[3][][]{\href{http://adsabs.harvard.edu/abs/#3}%
  {\def\hyper@linkstart##1##2{}%
    \let\hyper@linkend\@empty\citep[#1][#2]{#3}}}
\newcommandtwoopt{\citetads}[3][][]{\href{http://adsabs.harvard.edu/abs/#3}%
  {\def\hyper@linkstart##1##2{}%
    \let\hyper@linkend\@empty\citet[#1][#2]{#3}}}
\newcommandtwoopt{\citeyearads}[3][][]%
                 {\href{http://adsabs.harvard.edu/abs/#3}
                   {\def\hyper@linkstart##1##2{}%
                     \let\hyper@linkend\@empty\citeyear[#1][#2]{#3}}}
                 \makeatother
\usepackage{graphicx}                      \usepackage[hypcap]{caption}
\usepackage{subcaption}         
\usepackage{txfonts}
\begin{document} 

   \title{First  detection of  gas-phase ammonia  in  a planet-forming
     disk}

   \subtitle{NH$_3$, N$_2$H$^+$ and H$_2$O in the disk around TW Hya}

   \author{Vachail      N.      Salinas\inst{1}      \and      
     Michiel R.   Hogerheijde\inst{1}  \and   Edwin  A.   Bergin\inst{2}  \and
     L.  Ilsedore Cleeves\inst{3}  \and Christian  Brinch\inst{4} \and
     Geoffrey A. Blake\inst{5} \and Dariusz C. Lis\inst{6,7} \and Gary
     J.   Melnick\inst{3}  \and   Olja  Pani\'{c}\inst{8}   \and  John
     C.  Pearson\inst{9}   \and  Lars  Kristensen\inst{3}   \and  Umut
     A.  Y{\i}ld{\i}z\inst{9}   \and  Ewine  F.   van  Dishoeck\inst{1,10}
   }

   \institute{  Leiden Observatory,  Leiden University,  PO  Box 9513,
     2300                RA,                Leiden,                The
     Netherlands.\\ \email{salinas@strw.leidenuniv.nl} \and Department
     of Astronomy,  University of  Michigan, Ann Arbor,MI  48109, USA.
     \and   Harvard-Smithsonian  Center  for   Astrophysics,60  Garden
     Street, Cambridge, MA 02138,  USA.  \and Niels Bohr International
     Academy,   Niels  Bohr   Institute,   Blegdamsvej  17,   DK-2100,
     Copenhagen \O, Denmark \and  Division of Geological and Planetary
     Sciences, California Institute of Technology, Pasadena,California
     91125,  USA.  \and  LERMA,  Observatoire de  Paris, PSL  Research
     University,  CNRS, Sorbonne Universit\'es,  UPMC Univ.  Paris 06,
     F-75014,  Paris, France.   \and Cahill  Center for  Astronomy and
     Astrophysics   301-17,   California   Institute  of   Technology,
     Pasadena, CA 91125, USA.   \and Institute of Astronomy, Madingley
     Road, Cambridge,  CB3 0HA,  UK.  \and Jet  Propulsion Laboratory,
     California  Institute  of Technology,  Pasadena,  CA 91109,  USA.
     \and Max-Planck-Institut f\"{u}r Extraterrestrische Physik, 85748
     Garching, Germany.  
   } 
     \date{Received ; accepted }

  \abstract
{Nitrogen chemistry in protoplanetary disks and the freeze-out on dust
  particles is key to  understand the formation of nitrogen bearing
  species in early solar system  analogs.  In dense cores, 10\% to 20\%
  of the nitrogen reservoir is locked up in ices like NH$_3$, NH$_4^+$
  and  OCN$^-$. So  far,  ammonia  has not  been  detected beyond  the
  snowline in protoplanetary disks.}
{We  aim to  find  gas-phase  ammonia in  a  protoplanetary disk  and
  characterize its abundance with respect to water vapor.}
{Using HIFI  on the \emph{Herschel}  Space Observatory we  detect, for
  the first time, the ground-state rotational emission of ortho-NH$_3$
  in a protoplanetary disk, around  TW Hya.  We use detailed models of
  the disk's physical structure and the chemistry of ammonia and water
  to infer  the amounts of  gas-phase molecules of these  species.  We
  explore  two  radial distributions  (extended  across  the disk  and
  confined  to $<$60  au  like the  millimeter-sized  grains) and  two
  vertical  distributions  (near  the  midplane  and  at  intermediate
  heights above  the midplane, where  water is expected  to photodesorb
  off  icy   grains)  to  describe  the  (unknown)   location  of  the
  molecules. These  distributions capture the effects  of radial drift
  and vertical settling of ice-covered grains.}
{The  NH$_3$  $1_{0}$--$0_{0}$ line  is  detected simultaneously  with
  H$_2$O $1_{10}$--$1_{01}$  at an antenna  temperature of 15.3  mK in
  the  \emph{Herschel}  beam;  the  same spectrum  also  contains  the
  N$_2$H$^+$  6--5  line   with  a  strength  of  18.1   mK.   We  use
  physical-chemical models to reproduce  the fluxes with assuming that
  water and ammonia are  co-spatial. We infer ammonia gas-phase masses
  of 0.7-11.0  $\times$10$^{21}$ g,  depending on the  adopted spatial
  distribution,  in  line  with  previous literature  estimates.   For
  water, we  infer gas-phase  masses of 0.2-16.0  $\times$10$^{22}$ g,
  improving  upon  earlier literature  estimates  This corresponds  to
  NH$_3$/H$_2$O abundance ratios of  7\%-84\%, assuming that water and
  ammonia  are  co-located.   The  inferred  N$_2$H$^+$  gas  mass  of
  4.9$\times 10^{21}$ g agrees  well with earlier literature estimates
  based on  lower excitation  transitions. This masses correspond  to a
  disk-averaged     abundances    of     0.2--17.0$\times    10^{-11}$,
  0.1--9.0$\times 10^{-10}$ and 7.6$\times 10^{-11}$ for NH$_3$, H$_2$O
  and N$_2$H$^+$ respectively.}
{Only in  the most  compact and settled  adopted configuration  is the
  inferred NH$_3$/H$_2$O  consistent with interstellar  ices and solar
  system bodies  of $\sim$  5\%--10\%; all other  spatial distributions
  require  addition gas-phase  NH$_3$ production  mechanisms. Volatile
  release in the midplane may  occur via collisions between icy bodies
  if the available surface  for subsequent freeze-out is significantly
  reduced,  e.g.,  through growth  of  small  grains  into pebbles  or
  larger.}
   

   \maketitle
%

\section{Introduction}

The main  reservoir of  nitrogen-bearing species in  most solar-system
bodies is  unknown. The dominant form  of nitrogen on  these bodies is
inherited  from the  chemical  composition of  the  solar nebula  when
planetesimals   were   formed   \citep{Schwarz2014,Mumma2011}.    This
composition  depends strongly  on  the initial  abundances, which  are
difficult to probe since N  and $\rm{N_2}$ are not directly observable
in the interstellar medium (ISM). The Spitzer `Cores to Disks' program
found that  on average 10\% to  20\% of nitrogen is  contained in ices
like $\rm{NH_3}$, $\rm{NH_4^+}$, and $\rm{XCN}$, mostly in the form of
OCN$^-$  \citep{Oberg2011}. Water  is  the most  abundant volatile  in
interstellar ices  and cometary ices.   The relative abundance  of the
main nitrogen-bearing species compared to  water are of the order of a
few percent;  $\sim$ 5\% for  ammonia and $\sim$ 0.3\%  for $\rm{XCN}$
\citep{Bottinelli2010,Oberg2011}.   CN and HCN  have been  detected in
later   stages   of  star   formation   toward  protoplanetary   disks
\citep[see][]{Dutrey1997,Thi2004,Oberg2011b,Guilloteau2013} along with
resolved N$_2$H$^+$  emission in TW Hya  \citep{Qi2013} and unresolved
N$_2$H$^+$       emission      in       several       other      disks
\citep{Dutrey2007,Oberg2010,Oberg2011b}.   Although some  upper limits
exist  for  NH$_3$  in   protoplanetary  disks  in  the  near-infrared
\citep{Salyk2011,Mandell2012},  there are  no published  detections at
the moment.

Here we report the first detection of NH$_3$ along with the N$_2$H$^+$
6--5  line in the  planet-forming disk  around TW  Hya using  the HIFI
instrument  on the  \emph{Herschel} Space  Observatory. This  disk has
already been  well studied.  It was  first imaged by  the Hubble Space
Telescope  (HST) \citep{Krist2000,Weinberger2002}  revealing  a nearly
face-on   orientation.  \citet{Roberge2005}   took   new  HST   images
confirming  this orientation and  measured scattered  light up  to 280
au. Submillimeter  interferometric CO  data suggest an  inclination of
$6^\circ$ to $7^\circ$ \citep{Qi2004,Rosenfeld2012}. The age of TW Hya
is          estimated         to         be          8--10         Myr
\citep{Hoff1998,Webb1999,delaReza2006,Debes2013}  at a distance  of 54
$\pm$ 6 pc \citep{Rucinski1983,Wichmann1998,vanLeeuwen2007}.

This paper attempts  to model the ammonia emission  coming from TW Hya
assuming that  it is desorbed simultaneously with  water.  The thermal
desorption characteristics  of ammonia are  similar to those  of water
\citep{Collings2004}.   The  non-thermal  desorption  of  ammonia  via
photodesorption has a similiar rate  to that of water, within a factor
of three \citep{Oberg2009}. Ammonia is frozen in water-rich ice layers
present on interstellar dust particles.  Therefore, we can expect both
molecules to be absent from  the gas phase within similar regions.  In
order to properly constrain the NH$_3$/H$_2$O ratio we need to revisit
past models of water emission in the disk surrounding TW Hya.

The  ground-state  rotational emission  for  both  of  the water  spin
isomers has been found  around TW Hya by \citet{Hogerheijde2011} (from
now on H11),  also using the HIFI instrument  on \emph{Herschel} Space
Observatory.  They  explained this  emission using the  physical model
from  \citet{Thi2010} to  calculate the  amount of  water that  can be
produced by  photodesorption from a  hidden reservoir of water  in the
form of ice  on dust grains \citep{Bergin2010,vanDishoeck2014}.  Their
model overestimates the  total line flux by a  factor of {3--5.}  They
explore different ways to reduce the amount of water flux and conclude
that settling  of large icy grains is  the only viable way  to fit the
data.

Here, we  re-derive estimates of the  amount of water  vapor, using an
updated estimate of the disk gas  mass and considering the effect of a
more  compact distribution  of millimeter-size  grains, due  to radial
drift, as well as settling.  These dust processes are relevant for the
molecular  abundance of water  because they  can potentially  move the
bulk of the  ice reservoir away from regions  where photodesorption is
effective.  Simultaneously, we estimate the amount of NH$_3$ using the
detection  of  ammonia  in  the \emph{Herschel}  spectra,  and  derive
constraints on the NH$_3$/H$_2$O ratio  in the disk gas, assuming that
NH$_3$  and H$_2$O  are co-spatial.   We also  estimate the  amount of
N$_2$H$^+$  and compare  it to  the amount  of NH$_3$  using  a simple
parametric model.   Section \ref{sec:obs} presents our  data and their
reduction.  Section  \ref{sec:mod} contains our  modeling approach and
Section \ref{sec:res}  the resulting  ammonia and water  vapor masses.
Section  \ref{sec:dis}  discusses  the  validity  of  our  models  and
compares  these  predictions  to  standard  values.   Finally  section
\ref{sec:con} summarizes our conclusions.


\section{Observations}
\label{sec:obs}

\begin{figure}
\centering
\begin{subfigure}{0.98\columnwidth}
        \centering
        \includegraphics[width=\columnwidth]{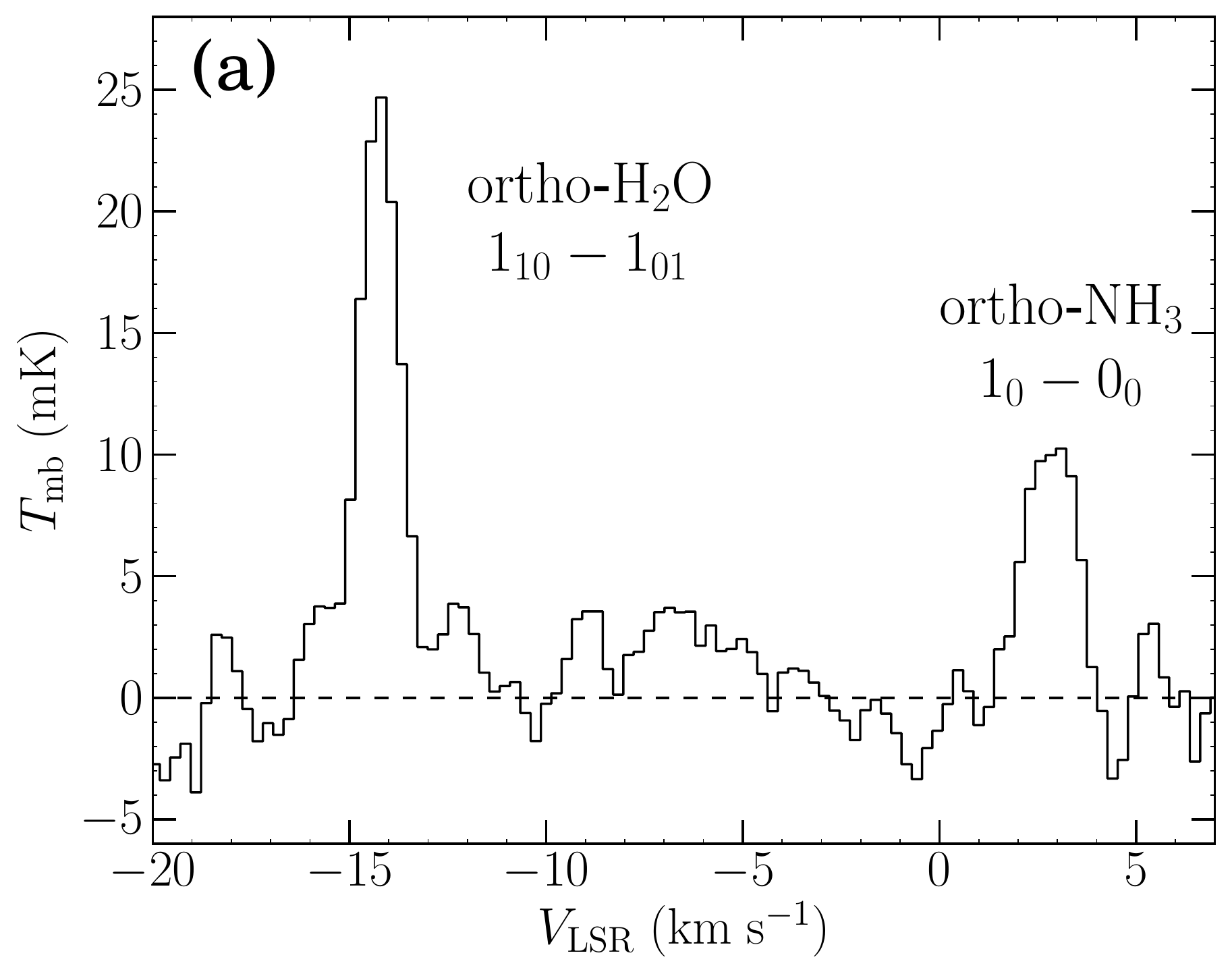}
    \end{subfigure}
    \hfill
    \begin{subfigure}{0.98\columnwidth}
        \centering
        \includegraphics[width=\columnwidth]{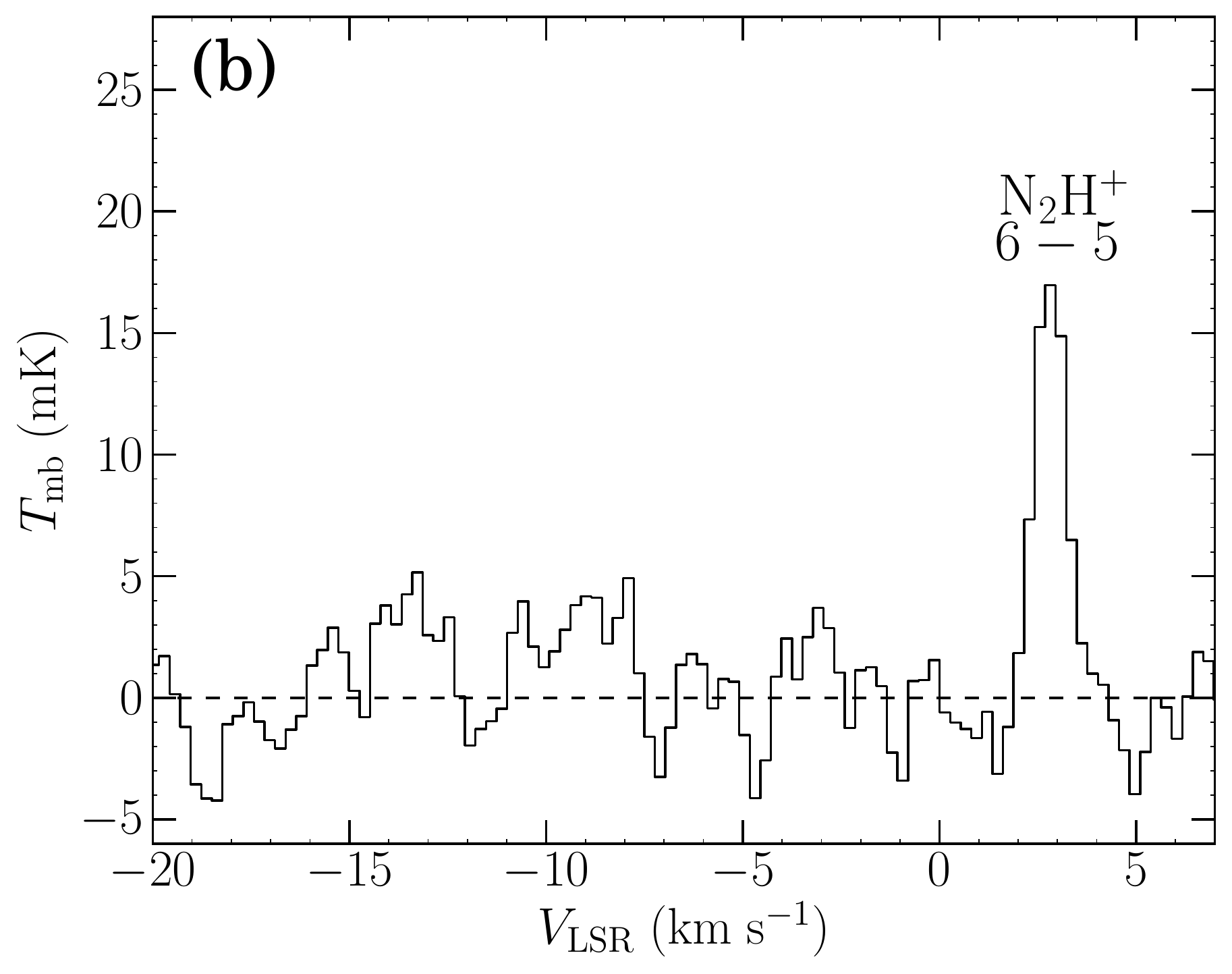}
    \end{subfigure} 
	\caption{Observed  spectra   of  (a)  ortho-NH$_3$  $1_0-0_0$,
          ortho-H$_2$O $1_{10}-1_{01}$ (earlier presented by H11), and
          (b) N$_2$H$^+$  6--5, using \emph{Herschel}  WBS. The dashed
          line shows the  continuum-subtracted spectral baseline.  The
          o-H$_2$O  and  o-NH$_3$   lines  are  observed  in  opposite
          sidebands, causing  the o-H${_2}$O to show up  at a velocity
          of $-$14 km s$^{-1}$ in panel (a).}
  \label{fig:spectra} 
\end{figure} 

Observations    of     TW    Hya    ($\rm\alpha_{2000}$     =    ${\rm
  11^h01^m51{\fs}91}$,  $\delta_{2000}$  =  $-34^\circ42'17{\farcs}0$)
were  previously presented by  H11 and  obtained using  the Heterodyne
Instrument  for the  Far-Infrared  (HIFI)  as part  of  the ‘Water  in
Star-Forming   Regions  with   \emph{Herschel}’  (WISH)   key  program
\citep{vanDishoeck2011}.   We now present  observations taken  on 2010
June 15 of  the NH$_3$ $1_0-0_0$ line at  572.49817 GHz simultaneously
with  o-H$_2$O at 556.93607  GHz using  receiver band  1b and  a local
oscillator tuning of 551.895 GHz (OBS-ID 1342198337).  We also present
the  detection  of  N$_2$H$^+$  6--5  at 558.96651  GHz  in  the  same
spectrum.   With  a  total  on-source  integration  of  326  min,  the
observation was taken with system temperatures of $360$--$400$ K.  The
data were  recorded in the  Wide-Band Spectrometer (WBS)  which covers
4.4 GHz with 1.1 MHz resolution.  This corresponds to 0.59 km s$^{-1}$
at  572 GHz.   The  data  were also  recorded  in the  High-Resolution
Spectrometer (HRS)  which covers 230 MHz  at a resolution  of 0.25 MHz
resulting in 0.13 km s$^{-1}$  at the observed frequency of the NH$_3$
$1_0-0_0$ line. The  calibration procedure is identical to  the one of
H11, but employs an  updated beam efficiency of $\eta_{\rm{mb}}=0.635$
and a HPBW of $36{\farcs}1$
\footnote{HIFI-ICC-RP-2014-001 on

http://herschel.esac.esa.int/twiki/bin/view/Public/HifiCalibrationWeb
} increasing the reported water line fluxes by about 17\% with respect
to  the values  of  H11.  Table  \ref{tab:lines}  summarizes the  line
fluxes    of   ammonia,   water    and   N$_2$H$^+$    6--5.    Figure
\ref{fig:spectra}  shows  the  calibrated  spectra  of  ortho-ammonia,
ortho-water and N$_2$H$^+$ 6--5 lines.

 \begin{table*}
 \caption{\label{tab:lines}Observed  line  parameters.}  \centering  
\begin{tabular}{c c c c c} 
\hline\hline   
Transition  & $F_{\rm{line}}$  $\rm{(10^{-19}~  W  m^{-2})}$  \tablefootmark{a,b}  &
$V_{\rm  LSR}$ (${\rm  km\,s^{-1}}$) \tablefootmark{b}  &  FWHM (${\rm
  km\,s^{-1}}$)\tablefootmark{c}            &            $\rm{T_{mb}}$
$(\rm{mK})$\tablefootmark{c}  \\ \hline  NH$_3$ $1_{0}-0_{0}$  (HRS) &
1.1$\pm$0.13 &  3.0$\pm$0.06 &  0.9$\pm$0.06 & 15.3$\pm$3.6  \\ NH$_3$
$1_{0}-0_{0}$  (WBS) &  1.1$\pm$0.10 &  2.9$\pm$0.06 &  1.4$\pm$0.06 &
11.3$\pm$2.0   \\   N$_2$H$^+$  $$6--5$$   (WBS)   &  1.0$\pm$0.11   &
2.9$\pm$0.03 & 0.9$\pm$0.04 & 18.1$\pm$2.4 \\ o-H$_2$O $1_{10}-1_{01}$
(HRS)  & 1.8$\pm$0.11  &  2.8$\pm$0.02 &  0.9$\pm$0.02 &  30.7$\pm$3.7
\\  o-H$_2$O $1_{10}-1_{01}$  (WBS)  & 1.9$\pm$0.09  & 2.9$\pm$0.03  &
1.3$\pm$0.03  &  24.0$\pm$2.0  \\  p-H$_2$O  $1_{11}-0_{00}$  (HRS)  &
6.7$\pm$0.62 & 2.7$\pm$0.05 &  1.1$\pm$0.05 & 41.0$\pm$8.1 \\ p-H$_2$O
$1_{11}-0_{00}$ (WBS)  & 6.7$\pm$0.44 & 2.7$\pm$0.04  & 1.3$\pm$0.04 &
39.0$\pm$5.2 \\ \hline
 \end{tabular}
\tablefoot{ \tablefoottext{a}{The errors  listed are calculated taking
    the  random  errors due  to  noise only  and  do  not include  the
    calibration uncertainty,  estimated to be about 20\%  of the total
    flux;   the  sideband   ratio  has   an  uncertainty   of  3-4\%.}
  \tablefoottext{b}{$F_{\rm{line}}$   is  the  integrated   flux  from
    $V_{\rm    LSR}$=    +1.5     to    +4.1    ${\rm    km\,s^{-1}}$}
  \tablefoottext{c}{Results of a gaussian fit. Errors on $V_{\rm LSR}$
    and  FWHM are  formal fitting  errors  and much  smaller than  the
    spectral resolution of 0.26 ${\rm km\,s^{-1}}$.}  }
\end{table*}

\section{Modeling approach}
\label{sec:mod} Since NH$_3$ is intermixed with H$_2$O on interstellar
ices and it is thought to desorb simultaneously \citep{Oberg2009}, our
modeling approach focuses on deriving  a NH$_3$/H$_2$O ratio in the TW
Hya  disk  assuming that  the  NH$_3$  emission  comes from  the  same
location as  the H$_2$O emission.  We  adopt a physical  model for the
gas density  and temperature and  re-derive the amount of  water vapor
from literature  results (H11).  Once  we define our H$_2$O  model, we
use  it  to  model  NH$_3$  emission  by  adopting  the  same  spatial
distribution as the water but  scaling the overall abundance as a free
parameter.  We also  take into account the effect  of radial drift and
vertical  settling   of  dust   grains  on  our   abundance  profiles.
Additionally,  we model  the N$_2$H$^+$  6--5 emission  by  assuming a
constant abundance throughout the  disk where the temperature is below
the CO  freeze-out temperature  (17 K) following  \citet{Qi2013}.  The
total amount  of N$_2$H$^+$  in this model  is also a  free parameter.
The following sections describe the physical and chemical structure of
our models.

\subsection{Physical structure}
Recently, \citet{Cleeves2014}  used HD measurements \citep{Bergin2013}
constrain the total  gas mass for the disk of  TW Hya of $0.04\pm0.02$
M$_{\sun}$, two  times more  massive that the  model used by  H11.  We
will adopt the physical structure of their best-fit model defined by a
dust surface density profile of
\begin{equation}
\Sigma_d(R)  = 0.04~{\rm g~cm^{-2}}\left(\frac{R}{R_c}\right)^{-1}{\rm
  exp}\left[-\left(\frac{R}{R_c}\right)\right]{\rm,}
\end{equation} 
and a scale height for small grains (and gas) given by
\begin{equation}
H(R) = 15~{\rm au}~\left(\frac{R}{R_c}\right)^{0.3}{\rm,}
\end{equation}
where  the critical  radius  $R_c$ is  150  au.  We  also adopt  their
estimated temperature profile $T(R,z)$ calculated from the ultraviolet
radiation   field   throughout   the   disk  \citep[see   Appendix   A
  of][]{Cleeves2014}.   They  did  not  consider a  radial  separation
between large and  small grains because the small  grains dominate the
dust surface  area, which is  most important for the  chemistry.  Many
models for  the TW Hya SED  include an inner  hole of radius a  few au
depleted  of dust.   Slight  variations  of this  gap  size have  been
proposed           \citep{Calvet2002,Eisner2006,           Hughes2007,
  Ratzka2007,Arnold2012,Menu2014}.   However, the observations  in the
large \emph{Herschel}  beam are not sensitive  to these small  scales, and we
ignore the inner hole in our model.

\subsection{Chemical model} 

Gas temperatures throughout  the disk in previous models  and ours are
typically below 200~K, thus excluding high temperature gas-phase water
formation.   We consider  thermal evaporation  and  photodesorption by
ultraviolet radiation as  H$_2$O production mechanisms and ultraviolet
photo-dissociation  and  freeze-out  as  the only  H$_2$O  destruction
mechanisms    using   the    time--dependent    chemical   model    of
\citet{Cleeves2014}. Thermal desorption is only dominant in the most inner disk up to a few au. Most of the water in this chemical model is released to the gas-phase through photodesorption in the outer disc. Compared to the chemistry used in H11, this model
uses  a more  realistic water  grain chemistry  and  updated gas-phase
reactions.    Figure~\ref{fig:model:dis}   summarizes   the   physical
conditions of  the model  and the  location of the  bulk of  the water
vapor.   A significant  decrease in  the midplane  abundance  of water
vapor can be seen in comparison to  the model proposed by H11 due to a
lower rate of  cosmic ray (CR) driven water  formation.  Two layers of
water abundance  can be distinguished  in Fig.~\ref{fig:model:dis}(b).
The   upper  layer  is   the  product   of  gas-phase   chemistry  and
photodesorption whereas  the layer at larger radii  but smaller height
is dominated by photodesorption.

Since our observations are spatially unresolved and the disk is nearly
face  on, no  information  of  the spatial  location  of the  emitting
molecules can be retrieved directly from our spectrally resolved data.
The  following  sections  describe  two processes  (radial  drift  and
settling)  due to  grain growth  that affect  the radial  and vertical
configuration of  dust grains that in turn  determine the distribution
of the ices.  We consider  two scenarios for the vertical location and
two scenarios for  the radial location of the  molecules, resulting in
four different configurations.

\subsubsection{Vertical Settling} 

For the vertical  distribution, we consider two extreme  cases.  In the
first scenario (p), the vertical distribution of the ammonia and water
follows  that  found  by   the  location  of  water  released  through
photodesorption (i.e.,  in the upper and intermediate  disk layers) as
described above. In the second extreme scenario (m) we assume that the
H$_2$O/H$_2$ and NH$_3$/H$_2$  abundances are constant.  We distribute
the  species vertically  out to  one  scale height  of the  millimeter
grains following \citet{Andrews2012},
\begin{equation}
	H(R)= 10.31\left(\frac{R}{\rm 100~ au}\right)^{1.25}{\rm au.}
\end{equation} Because the column density is dominated by the dense
layers near the midplane,  this model represents emission dominated by
the midplane (hence:  model m). The latter is  motivated by H11, where
they  tried to explain  water emission  coming from  TW Hya  using the
physical model  from \citet{Thi2010} to calculate the  amount of water
vapor  that  can be  produced  by  photodesorption.   But their  model
overestimates the total line flux  by a factor of 3-5.  They concluded
that settling of large(r) icy grains could be acting as a mechanism to
hide the icy grains from the reach of ultraviolet photons resulting in
the  lower-than-expected  water  line  fluxes.   We do  not  make  any
assumptions about  the production  mechanism of the  gas-phase ammonia
and water in  the absence of photodesorption in  the midplane (m), but
discuss possible mechanisms in section \ref{sec:dis}.

\subsubsection{Radial drift} 

For the radial location we consider an extended model (E) with ammonia
and water across  the entire disk out to 196  au, corresponding to the
extent of $\mu$-size grains\citep{Debes2013},  and a compact model (C)
with  NH$_3$ and  H$_2$O confined  to the  location of  the millimeter
grains ($<$60 au).  \citet{Andrews2012} find that the millimeter-sized
grains are located within 60 au,  likely as the result of radial drift
causing a separation between the large-size population of dust and the
small-size population  of dust which  remains coupled to the  gas. The
compact  model (C)  is also  motivated by  H11, since  grains settling
operates faster  than drift because the vertical  pressure gradient is
larger than the radial one.  Any grains large enough to drift radially
will certainly have settled vertically first. This means the molecules
are already locked up in large(r) grains when they experience (or not)
a radial  drift.  Our  compact model (C)  represents the  extreme case
where all water and ammonia ice  has been transported to within 60 au,
and is (partially) returned to the  gas phase only there.  In the same
way, our  extended model  (E) represents the  other extreme  where the
water and ammonia reservoir, locked  up in icy dust particles, extends
across the full disk.

\begin{figure} 
\centering
\begin{subfigure}{0.98\columnwidth}
        \centering
        \includegraphics[width=\columnwidth]{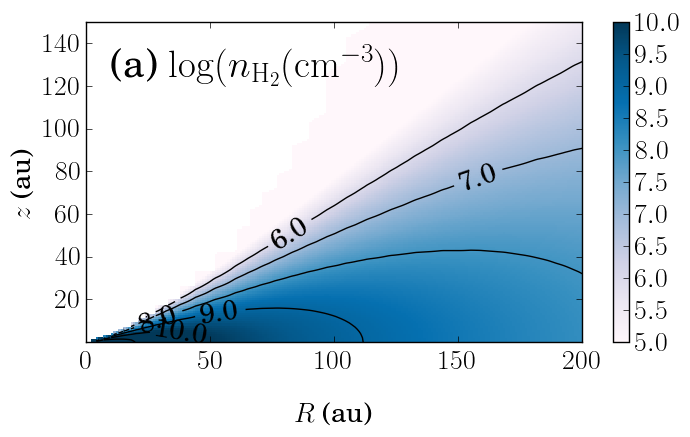}
    \end{subfigure}
    \hfill
    \begin{subfigure}{0.98\columnwidth}
        \centering
        \includegraphics[width=\columnwidth]{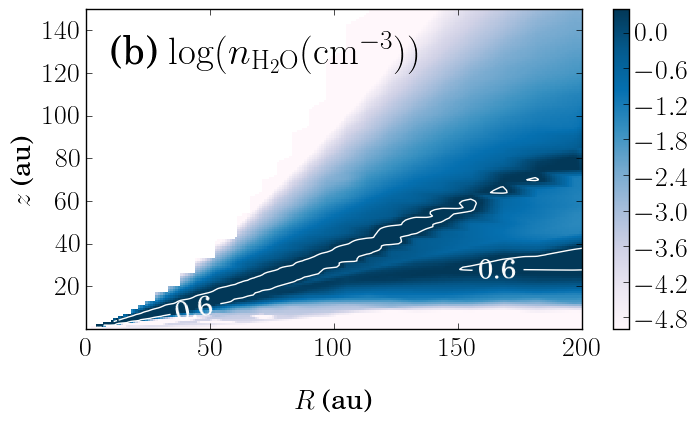}
    \end{subfigure}
    \hfill
     \begin{subfigure}{0.98\columnwidth}
        \centering
        \includegraphics[width=\columnwidth]{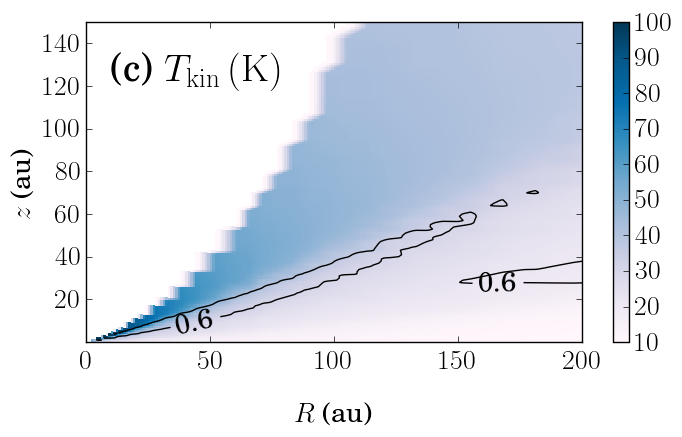}
    \end{subfigure}
    \caption{ Overview of the  adopted model structure.  (a) logarithm
      of the  molecular hydrogen density in  $\rm{cm}^{-3}$ with black
      contours.  (b)  logarithm of the water vapor  density with white
      contours at 0.6 $\rm{cm}^{-3}$.  (c) shows the gas temperature,
      in  K.  Black  contours  represent the  logarithm  of the  water
      abundance at 0.6 $\rm{cm}^{-3}$.}
    \label{fig:model:dis}
 \end{figure} 

\begin{figure}
\includegraphics[width=.98\columnwidth]{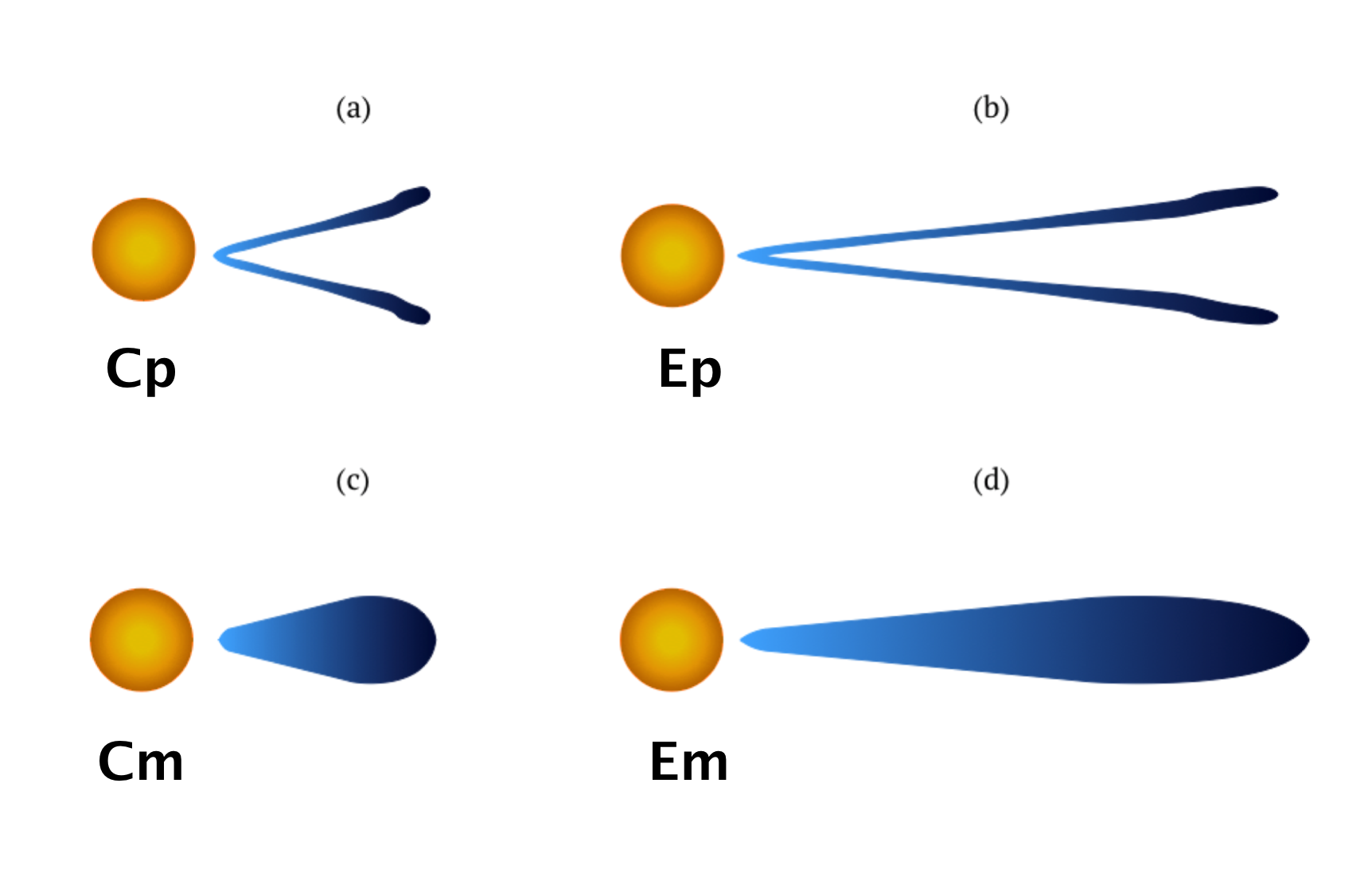}
\caption{Schematic illustration of the  location of water vapor in our
  models:  (a) Compact  photodesorption-layer configuration  (Cp), (b)
  Extended photodesorption-layer configuration  (Ep), (c) Compact with
  a  constant abundance configuration  (Cm), (d)  and Extended  with a
  constant abundance configuration (Em).}
\label{fig:models}
\end{figure}

As Fig.  \ref{fig:model:dis} shows, water vapor is mostly present in a
thin photo-dominated layer (models p)  or near the midplane (models m)
following  total  H$_2$  density  profile.   Figure  ~\ref{fig:models}
summarizes the resulting four different scenarios (Em, Ep, Cm, Cp). In
all scenarios, the  total amount of ammonia and water  vapor is a free
parameter  constrained  from fitting  the  observed  line fluxes.   In
particular,  for  the p-models,  this  means  we  use the  radial  and
vertical density distribution from the detailed calculations but scale
the total amount of ammonia and water vapor up or down as necessary.

\subsection{Line excitation and radiative transfer}\label{sec:lime}

We  use   LIME  (v1.3.1),  a   non-LTE  3D  radiative   transfer  code
\citep{Brinch2010} that can predict  line and continuum radiation from
a source model. All of our  models use 15000 grid points. Doubling the
number of grid points does not affect the outcome of the calculations.
Grid points are distributed randomly in $R$ using a logarithmic scale.
This means  in practice that  inner regions of  the disk have  a finer
sampling than the  outer parts of the disk.  Since  it is difficult to
establish reliable convergence criteria, LIME requires to manually set
the number  of iterations  of each  point.  We set  this number  to 12
confirming  convergence  in   our  models  by  performing  consecutive
iterations.  Forty  channels of 110 m  s$^{-1}$ each are  used for all
line models  with 200 pixels of  0.05 arcsec.  Since  we are comparing
these models  with spatially unresolved  data, we calculate  the total
flux by summing all the pixels after subtracting the continuum.

Rate coefficients for ortho-ammonia,  N$_2$H$^+$ and both spin isomers
of  water are  taken from  the  Leiden Atomic  and Molecular  Database
\citep{LAMBDA_database2005}\footnote{www.strw.leidenuniv.nl/~moldata/}.
The excitation  levels of  para-H$_2$O and ortho-H$_2$O  have separate
coefficients for o-H$_2$ and p-H$_2$ \citep{Daniel2011}.  The o-NH$_3$
collision rates  are only  available for p-H$_2$,  so we  regard total
H$_2$    as   p-H$_2$   to    calculate   their    population   levels
\citep[see][]{Danby1988}. The  N$_2$H$^+$ collision rates  are adopted
from   HCO$^+$   following   \citet{Flower1999}.   We   assume   H$_2$
ortho-to-para ratios in local  thermal equilibrium. Given the low dust
temperatures this  implies H$_2$ OPR  $<0.3$.  If instead  we increase
the H$_2$ ortho--to--para  ratio to 3, the high  temperature limit for
formation on  grains \citep{Flower1996},  it will increase  the H$_2$O
line fluxes increase  by a factor $\sim 2$.  We  discuss the effect of
this on the inferred water vapor mass below.

\section{Results}
\label{sec:res}

\begin{figure}[!htb] 
\centering
\begin{subfigure}{0.98\columnwidth}
        \centering
        \includegraphics[width=\columnwidth]{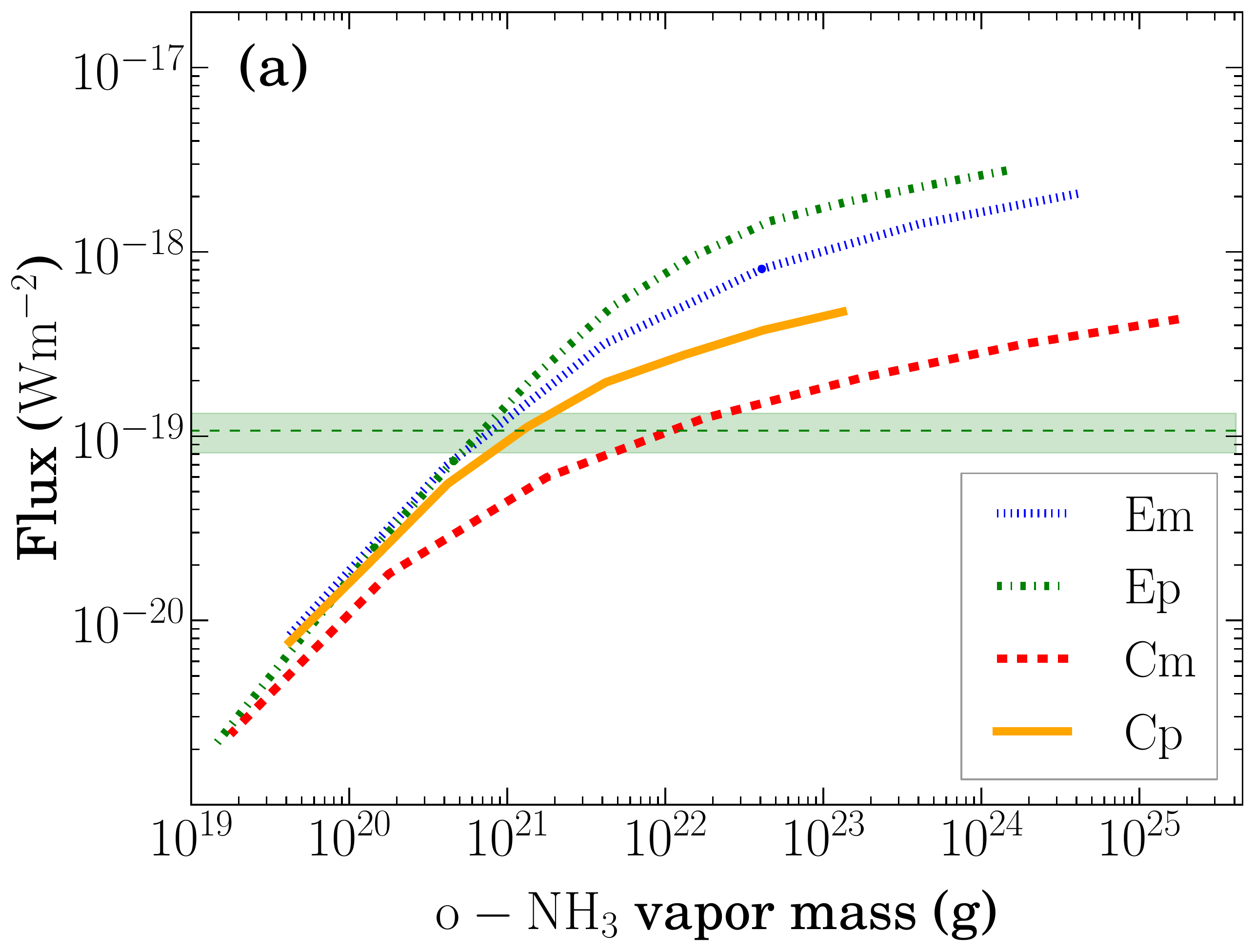}
    \end{subfigure}
    \hfill
    \begin{subfigure}{0.98\columnwidth}
        \centering
        \includegraphics[width=\columnwidth]{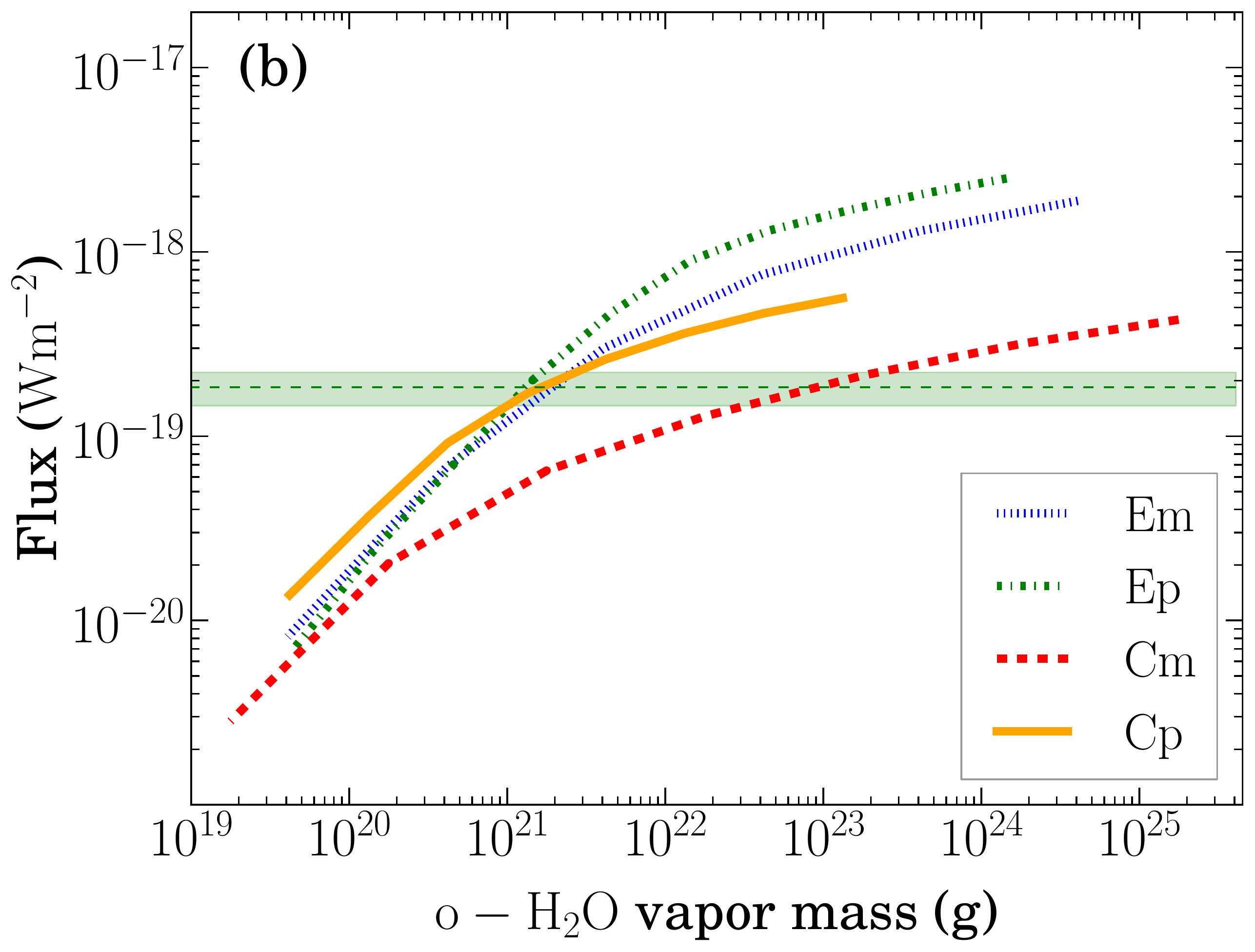}
    \end{subfigure}
     \begin{subfigure}{0.98\columnwidth}
        \centering
        \includegraphics[width=\columnwidth]{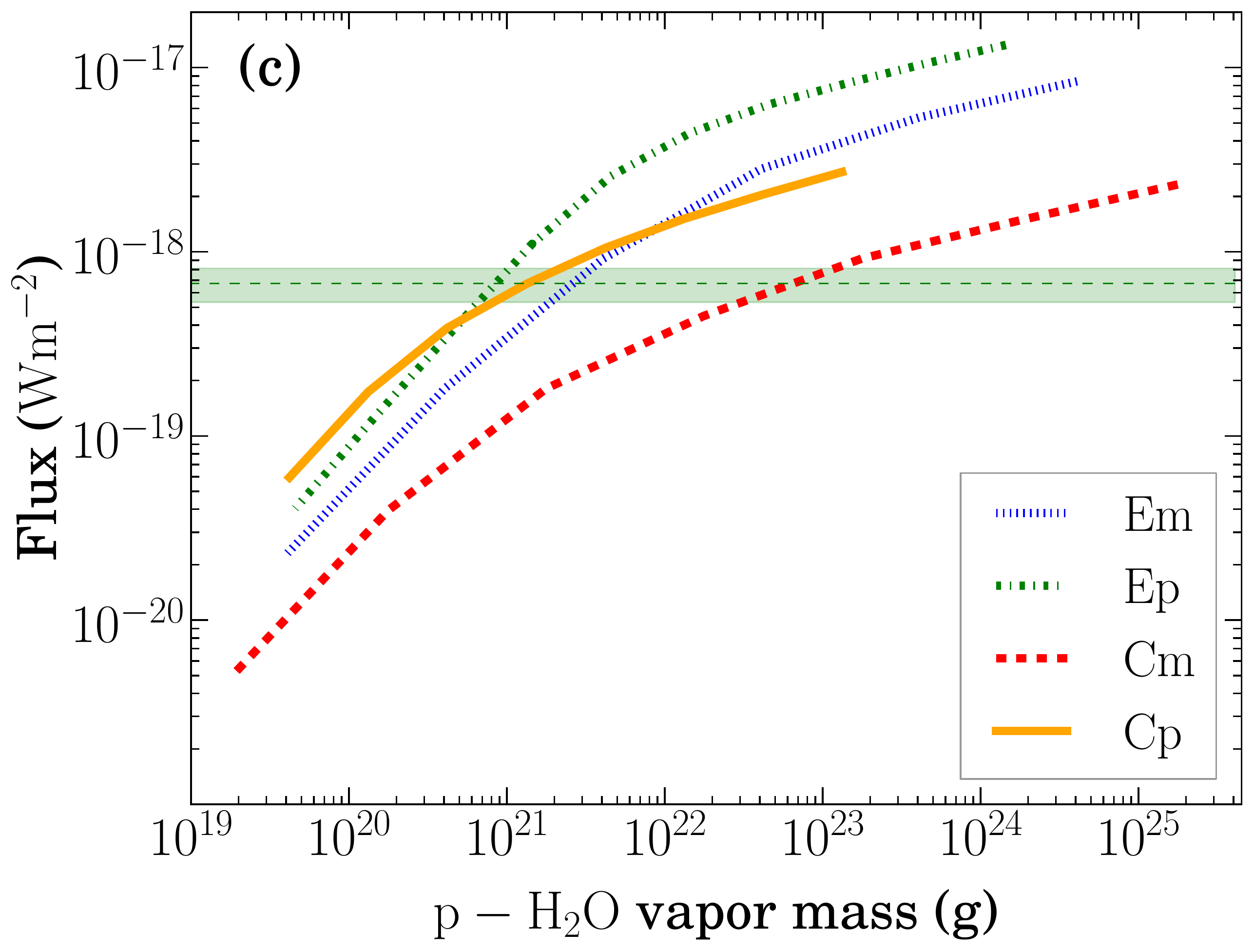}
    \end{subfigure}
    \caption{ Figures  (a), (b) and (c)  shows LIME \citep{Brinch2010}
      output   of  the   total  line   fluxes   of  ortho-$\rm{NH_3}$,
      ortho-$\rm{H_2O}$   and  para-$\rm{H_2O}$   respectively   as  a
      function of  the total water  vapor mass.  The blue,  green, red
      and yellow  curves correspond  to the Em,  Ep, Cm and  Cp models
      respectively.  The  dashed lines  and horizontal green  bar show
      the observed line fluxes  and their 3$\sigma$ ranges, with sigma
      having two sources of  noise added in quadrature. The systematic
      error of  the observations,  estimated to be  about 20\%  of the
      total flux, and the rms of the spectra.}
    \label{fig:results}
 \end{figure}

Figure~\ref{fig:results}   shows  the  emerging   line  flux   in  the
\emph{Herschel}  beam   as  function   of  ammonia  and   water  vapor
mass.  These curves  of growth  (flux ($F$)  vs  mass~$\propto$ column
density($N$)) are consistent with  saturated lines: the slopes go from
linear  ($F\propto  N$)  in   the  low  opacity  regime  to  saturated
($F\propto$ ${\rm  \sqrt{{\rm ln}(N)}}$).  The latter  behavior is due
to the line becoming gradually optically thick in its wings, resulting
in a steady flux growth.

Our four  models predict different  asymptotic values for  large vapor
masses. In the asymptotic regimes lines are fully thick and probe only
a very thin region near the  surface of the disk.  The larger E models
therefore result in more flux compared  to the smaller C models. The m
models trace  higher H$_2$ densities  but lower temperatures  than the
corresponding p models, resulting  in different predicted fluxes. This
has a strong  effect on the water lines,  that have critical densities
and upper level  energies higher than the conditions  prevalent in the
regions where the lines originate. This effect is strongest for the Cm
models because  in this regime  densities are considerably  larger and
the lines becomes thermalized and opaque, resulting in larger required
water-vapor masses.

Table  \ref{tab:summary} summarizes the  best-fit vapor  masses; error
estimates  include  statistical errors  on  the  observations and  the
systematic errors on the total  line flux, estimated to be about 20\%.
The ammonia to water ratios shown in Table~\ref{tab:summary} assume an
OPR  of  ammonia  of  either  $\infty$  or  1.   As  stated  above  in
section~\ref{sec:lime} we assume H$_2$ ortho--to--para ratios in local
thermal  equilibrium.   If we  increase  this  value  to 3  (the  high
temperature limit), then the derived masses would decrease by a little
more than 1 order of magnitude in our most massive and optically thick
model (Cm)  and by less than  one order of magnitude  in the remaining
models (Cp,  Em, Ep).  We do not  include this in the  error budget of
our reported values because the NH$_3$ and N$_2$H$^+$ might be equally
affected.

All four  models yield o-NH$_3$ masses  ranging from (0.7--1.2)$\times
10^{21}$~g for  models Ep, Em, and Cp,  to 1.1$\times10^{22}$~g. These
correspond  to   ammonia  abundances  ranging   from  (2.0-9.5)$\times
10^{-12}$ (E models) to (0.6-1.7)$\times10^{-10}$ (C models), relative
to H$_2$.   For water, a higher  range of masses  is inferred, ranging
from  (2.2--4.5)$\times  10^{21}$~g for  models  Ep,  Em,  and Cp,  to
$1.6\times  10^{23}$  g  for  model  Cm.  These  correspond  to  water
abundances  ranging  from  (1.1-3.0)$\times  10^{-11}$ (E  models)  to
(4.5-9.0)$\times10^{-10}$  (C models), relative  to H$_2$.   The water
OPR is found to range from  0.73 to 1.52. If the associated errors are
considered then  the range is much  wider (0.2-3.0) with  model Ep, Cp
and  Cm in  agreement  with  the interstellar  and  cometary range  of
2.0-3.0 for the  OPR of water. The ammonia to  water ratio ranges from
7\%$^{9\%}_{-5\%}$ to 84\%$^{61\%}_{-36\%}$.

Calculations     using     a     simple    escape-probability     code
\citep[RADEX,][]{vanderTak2007}  reproduce  the  observed line  fluxes
adopting   the  inferred   vapor  masses   and  using   densities  and
temperatures representative for the emitting regions.  But only a full
3D  calculation can  reproduce the  exact line  fluxes because  of the
large range in densities  and temperatures.  These simple calculations
also show that, under the conditions of the four models, equal amounts
of  o--$\rm{H_2O}$ and  o--$\rm{NH_3}$ give  approximately  equal line
strengths (within 50\%).   This means that we can  relate the observed
line ratios of $F_{\rm{o-NH_3}}$/$F_{\rm{o-H_2O}}\sim$ 0.6 to estimate
the  actual  overall  $\rm{NH_3/H_2O}$  abundance  fraction  of  about
0.35-0.65 as confirmed by the  detailed LIME modeling below. The large
NH$_3$/H$_2$O ratios  suggested by most models are  therefore a direct
consequence  of  the near-equal  observed  line  fluxes  of H$_2$O  and
NH$_3$; only  in the Cm model  where lines are opaque,  are much lower
NH$_3$/H$_2$O values consistent with the observed fluxes.

\def\arraystretch{1.2}
\begin{table*}
  \caption{\label{tab:summary}Summary   table   of   our  results   in
    different regimes.}  \centering 
  \begin{tabular}{c c c c c} 
\hline\hline  
~  & Ep  & Em&  Cp & Cm\\  \hline   Total   $\rm{o-H_2O}$   vapor  mass   (10$^{21}$g)   &
1.3$^{+0.4}_{-0.3}$  &  1.9$^{+0.6}_{-0.5}$  &  1.6$^{+1.1}_{-0.6}$  &
94$^{+111.2}_{-5.8}$ \\ Total  $\rm{p-H_2O}$ vapor mass (10$^{21}$g) &
0.9$^{+0.2}_{-0.2}$  &  2.6$^{+0.8}_{-0.7}$  &  1.4$^{+0.7}_{-0.6}$  &
65$^{+5.4}_{-3.4}$  \\  Total $\rm{o-NH_3}$  vapor mass  (10$^{20}$g) &
7.0$^{+1.8}_{-1.9}$   &   8.0$^{+0.3}_{-0.3}$   &   12$^{+7}_{-4}$   &
110$^{+130}_{-60}$     \\   OPR     &     1.52$^{+0.58}_{-0.49}$    &
0.73$^{+0.32}_{-0.28}$        &        1.14$^{+0.97}_{-0.65}$        &
1.38$^{+2.07}_{-1.15}$  \\ $\rm{NH_3}$/$\rm{H_2O}$\tablefootmark{a} &
33\%$^{+11\%}_{-11\%}$       $\sim$       66\%$^{21\%}_{-22\%}$      &
19\%$^{8\%}_{-7\%}$        $\sim$        38\%$^{17\%}_{-15\%}$       &
42\%$^{30\%}_{-18\%}$       $\sim$       84\%$^{61\%}_{-36\%}$       &
7\%$^{9\%}_{-5\%}$ $\sim$ 15\%$^{20\%}_{-11\%}$ \\ \hline
  \end{tabular} \tablefoot{The errors on the masses include noise and
20\%  calibration error.   \tablefoottext{a}{The values  correspond to
  the number  ratio, not the mass  ratio, and are  calculated with two
  extreme  OPR  of ammonia  $\infty$  and  1.   Errors are  calculated
  propagating  the   noise  and   calibration  error  of   each  total
  abundance.}  }
\end{table*}

Our  best-fit model  result for  N$_2$H$^+$ is  in agreement  with the
N$_2$H$^+$ 4--3  emission reported in  \citet{Qi2013}. The fit  to the
N$_2$H$^+$  6--5 emission  yields  a total  N$_2$H$^+$  vapor mass  of
4.9$\times   10^{21}$~g,   $\sim$50\%  higher   than   the  model   of
\citet{Qi2013} obtained from integrating their best-fit model ($z_{\rm
  big}(H)=3$ in table S2).

\section{Discussion}
\label{sec:dis}

\subsection{Ice reservoirs and total gass masses}\label{sec:icy} 

The  inferred  NH$_3$  vapor  masses  from  7.0$\times  10^{20}$~g  to
1.1$\times  10^{22}$~g are  much  smaller than  the potential  ammonia
\emph{ice} reservoir of $\sim 3.0\times10^{28}$ g.  This ice reservoir
mass estimate  was obtained  assuming an elemental  nitrogen abundance
relative  to  H  of   $\rm{2.25\times10^{-5}}$  and  a  disk  mass  of
0.04$\pm$0.02 M$_{\sun}$,  and assuming that 10\%  of nitrogen freezes
out  on grains  as  NH$_3$  \citep{Oberg2011}.  In  the  same way,  we
estimate a water ice  mass reservoir of $3.4\times10^{29}$ g, adopting
an oxygen  elemental abundance of  $\rm{3.5\times10^{-4}}$ relative to
H, assuming  that 70\% of O  is locked up  in water \citep{Visser2009}
and all  of it is  frozen out. Both  mass estimates indicate  that the
detected vapor  masses are only tiny  fractions ($\lesssim10^{-6}$) of
the available ice reservoirs.

The   total    water   mass   of   our    original   chemistry   model
($4.6\times10^{24}$g)  is  2.5  to   25  times  (Cm  and  Ep  models
respectively) more massive than the values derived from our reanalysis
of the water detection towards TW  Hya.  H11 reported a model 16 times
more  massive than  their original  chemistry model  to fit  the water
emission coming from  the disk of TW Hya.  This value is significantly
lower than the value of our analogous Ep model which indicates an even
higher degree of settling of  the icy grains than previously proposed.
This is  consistent with earlier  conclusions that most  volatiles are
locked     up     in      big     grains     near     the     midplane
\citep{Hogerheijde2011,Du2015}. 

\subsection{Gas phase chemistry}\label{sec:chem}

The  reported ammonia  to water  ratios are  considerably  higher than
those found for  ices in solar system comets  and interstellar sources
\citep{Mumma2011}, which  are typically  below 5\%. Our  model derived
ratios assume that the NH$_3$  and H$_2$O emission originates from the
same  regions;  however, if  this  is  not  the case,  expressing  the
relative amounts of  ammonia and water as a ratio  is not very useful.
Then, it  is better to work  with their (disk  averaged) abundances of
0.2-17.0$\times$10$^{-11}$ and 0.1-9.0$\times$10$^{-10}$ respectively.

An obvious  conclusion from the large  amount of NH$_3$  is that other
routes   exist  for  gas-phase-NH$_3$   in  addition   to  evaporating
NH$_3$/H$_2$O ice mixtures.  In the colder outer disk the synthesis of
ammonia in the gas phase  relies on ion-molecule chemistry. This means
that N$_2$ needs to be broken apart (to release N or N$^+$) first, but
N$_2$  can self-shield  against photodissociation  \citep{Li2013}. The
chemistry in  the disk of  TW Hya seems  to reflect an  elevated X-ray
state  of  the  star  \citep{Cleeves2014}.  This  strong  X-ray  field
scenario  could  be  invoked  to  break  N$_2$  apart.  The  models  of
\citet{Schwarz2014}  for a  typical T  Tauri  disk (with  a FUV  field
measured in TW Hya) give values for the abundance of NH$_3$ as high as
$\sim$10$^{-8}$   \cite[the   models   of][also   produce   the   same
  abundance]{Walsh2015},  which  would be  sufficient  to explain  the
emission.  Modeling  tailored to TW  Hya with the correct  stellar and
disk parameters, as well as appropriate initial conditions is required
to  fully address  the  question of  the  origin of  the large  NH$_3$
abundance.

Of our  four models, the  Cm model stands  out in that it  yields much
lower  NH$_3$/H$_2$O ratios that  are consistent  with the  low values
found in  solar system  bodies and interstellar  ices. It is  also the
only model  where the (large  uncertainties of the) derived  water OPR
overlap   with  the  2-3   range  commonly   found  in   solar  system
comets. Recent laboratory work by \citep{Hama2015} shows that water in
ices efficiently attains an OPR  of 3 upon release into the gas-phase,
indicating  that  the  OPR  is   not  a  reflection  of  the  physical
temperature  and that  high OPR  values are  naturally  expected.  The
NH$_3$/H$_2$O values and the water OPR values together can be taken as
evidence  that  the  Cm  model   is  a  correct  description  for  the
distribution of H$_2$O  and NH$_3$ in the disk. If  so, a mechanism to
release water in the midplane is required.

\subsection{Collisions of large bodies as a production mechanism}\label{sec:coll}

In the midplane models (Em and Cm), photodesorption cannot explain the
abundance  of water  and  ammonia  in the  gas  phase, as  ultraviolet
radiation cannot penetrate to these depths. CR-induced H2O desorption,
such as  modeled in  H11, cannot produce  the required amount  of H2O.
The typical  water vapor  abundances found in  the H11  chemical model
near  the midplane are  of order  $X_{\rm{H_2O}} \sim  10^{-13}$, much
smaller than the corresponding  best-fit midplane abundances in the Em
and  Cm  models of  $X_{\rm{H_2O}}  \sim  10^{-10}-10^{-9}$.  How  can
volatiles such as ammonia and  water exist near the midplane where low
temperatures and high densities would ensure rapid freeze out?
  
One way of releasing such a vapor mass from the icy reservoir would be
through collisions  of larger  icy bodies. We  can calculate  how much
water  needs to  be released  through these  collisions, if  we assume
steady  state  with freeze  out,  to  retain  the observed  amount  of
volatiles  in the  gas  phase.   Freeze out  is  calculated using  the
freeze-out   rate   expression   for   neutral  species   derived   in
\citet{Charnley2001}. For  typical temperatures of 12  K and densities
of 1.7$\times10^{9}$cm$^{-3}$,  the freeze out rate of  water vapor is
$\rm{2.5\times10^{13}g~s^{-1}}$ if we consider the (Em) model to match
the  observations.   That   is  equivalent  to  completely  destroying
$\sim$7,000 comets per year, with  a mass comparable to Halley's comet
and assuming they  consist of 50\% water. After 10  Myr roughly 5\% of
the water previously locked up in  icy grains would be back to the gas
phase if  an ice reservoir  of $\rm{3.4\times10^{29}g}$ is  present in
the disk. In  the case of the (Cm) model a  higher production of water
vapor   $\rm{1.6\times10^{16}g~s^{-1}}$  is   needed   to  match   the
observations.   This  would  mean destroying  $\sim\rm{5\times10^{6}}$
comets per  year. After 10 Myr  we would have produced  ten times more
water  vapor than  its total  ice  reservoir.  Such  large numbers  of
collisions  and  the  significant  (or  even  unrealistic)  amount  of
released  water suggest  that  collisions between  icy  bodies are  an
unlikely  explanation for  the observed  amount of  water  and ammonia
vapor in the midplane models.

The  freeze-out  rates used  above  have  been  calculated assuming  a
typical    grain   size    distribution    $n(a)\varpropto   a^{-3.5}$
\citep{Mathis1977} with minimum and  maximum grains sizes of 10$^{-8}$
m and 10$^{-1}$  m.  Since the majority of  the surface for freeze-out
is  on (sub)$\mu$m-size  grains,  we  can expect  this  surface to  be
substantially  reduced  if  these  smaller  grains  are  removed  thus
reducing  the  freeze-out rate  significantly.   Small  grains may  be
removed by photoevaporating  winds \citep{Gorti2015}, when transported
to the upper layers by vertical mixing, or have coagulated into larger
grains.  In the extreme case  where all of the $\mu$m-size grains have
grown into  larger bodies  the freeze-out rate  can be reduced  by two
orders of magnitude.

We  can get  a drop  by a  factor  of 100  in the  freeze-out rate  by
directly   calculating   the  mean   grain   surface   in  Eq.6   from
\citet{Charnley2001} by  setting $a_{\rm min}$=10$^{-4}$ m  for our Em
model,   $a_{\rm  min}$=10$^{-3}$m   for  our   Cm   model  \citep[see
  calculations of][]{Vasyunin2011}  and $a_{\rm max}$=10$^{-1}$  m for
both.   If this  assumption  holds  then our  model  with the  highest
production  rate (Cm)  will  have  processed only  10\%  of its  water
reservoir  into water  vapor  in the  span  of 10  Myr, equivalent  to
destroying  only $\sim$5000  comets per  year.  In  the same  way, the
amount of water  processed in the span  of 10 Myr in our  (Em) will be
only  0.5\%  of  its  ice  reservoir, equivalent  to  destroying  only
$\sim$70 comets per year.  Such number are much more realistic, making
this a viable mechanism to release volatiles in the midplane.

Nevertheless, for  this scenario  to be viable,  the system  must meet
some criteria. First,  the comets (or planitesimals) must  have a high
enough  collision  rate  that   accounts  for  the  numbers  estimated
above. In the  outer disc this can be  enhanced through shepherding by
planets,  i.e., sweeping  up the  planetesimals into  one proto-debris
belt.     \citet{Acke2012}    calculate    a   collision    rate    of
$\rm{6.3\times10^{13}g~s^{-1}}$   in   Fomalhout's   debris  disk   to
reproduce the thin dust belt  seen in far-infrared images (70-500 $\mu
m$).   This rate  is  comparable to  our  estimated rate  even in  the
absence of a reduction in  grain surface available for freeze out.  In
the  case of  Fomalhout,  this  rate corresponds  to  a population  of
$\rm{2.6\times10^{11}}$ 10-km-sized comets comparable to the number of
comets    in     the    Oort    Cloud     of    $\rm{10^{12}-10^{13}}$
\citep{Weissman1991}.   Second,  the  collisions must  release  enough
energy to sublimate  the ices.  This is only  achieved if the relative
velocities  of  the  parent  bodies  are sufficiently  high.   If  the
colliding  bodies have high  eccentricities their  relative velocities
can be large. But in the presence of gas, we expect their orbits to be
circularised.  If this is the  case, then the relative velocities will
be dominated  by the radial drift in  the outer parts of  the disk and
enhanced in the very inner  regions by turbulence.  Finally, the small
dust produced in the  cometary (or planitesimal) collisions themselves
must not provide a surface for the volatiles to freeze back onto.

If a sufficient number of small grains ($\lesssim \mu$m) is removed by
coagulation  into  larger  grains  ($\gtrsim$ mm)  and  the  relatives
velocities  and   rates  of  the  collisions   between  larger  bodies
($\gtrsim$  m)  in the  miplane  are  sufficiently  high to  meet  the
conditions above,  then collisions between  icy bodies is  a plausible
mechanism to release  (and keep in the gas phase  for long enough) the
amount of water  and ammonia that we observe.   The treatment above is
simplistic  and there  are  many  other ways  to  achieve this  (e.g.,
changing the slope of the grain distribution or by photoevaporation of
grains along with the gas). A full treatment of the combined effect of
grain growth,  drift, settling, collisions and  volatile freeze-out is
needed to confirm this scenario but is beyond the scope of this paper.

\section{Conclusions}
\label{sec:con}
 
We  have  successfully detected  NH$_3$  and  N$_2$H$^+$  in the  disk
surrounding  TW  Hya.   We  use  a non-LTE  excitation  and  radiative
transfer code and  a detailed physical and chemical  disk structure to
derive the  amount of NH$_3$, N$_2$H$^+$, and  (for comparison) H$_2$O
adopting four  different spatial distributions of  the molecules.  Our
main conclusions are as follows.

\begin{enumerate}

\item The  NH$_3$ emission corresponds  to an ammonia vapor  mass that
  ranges  from $7.0\times10^{20}$g  (Ep model)  to $1.1\times10^{22}$g
  (Cm model).
  
\item We  use the  above values  and the same  approach to  get H$_2$O
  vapor masses to derive NH$_3$/H$_2$O ratios ranging from 7\% to 15\%
  (Cm model)  and 42\% to  84\% (Cp model),  adopting a NH$_3$  OPR of
  $\infty$  or 1, respectively.   These ratios  are higher  than those
  observed in  solar system and interstellar ices,  with the exception
  of our most massive and compact configuration (Cm model).

\item Of our four models,  only model Cm gives NH$_3$/H$_2$O ratios as
  low as observed in interstellar  ices and solar system comets. It is
  also the  only model that, within  the errors, gives a  water OPR of
  2-3 comparable to solar system comets. This can be taken as evidence
  that H$_2$O  and NH$_3$ are indeed  located near the  midplane at at
  radii $<$60 au.

\item  If the  H$_2$O and  NH$_3$  follow the  Cp, Ep,  or Em  spatial
  distributions,  the  implied high  NH$_3$/H$_2$O  ratio requires  an
  additional  mechanism to  produce gas-phase  NH$_3$. A  strong X-ray
  field may provide  the necessary N atoms or  radicals to form NH$_3$
  in the gas.

\item If  NH$_3$ and  H$_2$O emission comes  from the  midplane, where
  photodesorption does  not operate  (models m), collisions  of larger
  bodies can release NH$_3$ and H$_2$O and explain the observed vapor.
  This requires a  reduction of the total grain  surface available for
  freeze  out, e.g.,  through the  growth of  grains into  pebbles and
  larger;  and a  sufficiently  high collision  rate and  sufficiently
  violent collisions to release the volatiles.

\item The ammonia vapor mass is similar to the N$_2$H$^+$ mass derived
  with our  simple model and  to that inferred by  \citet{Qi2013} (see
  section \ref{sec:res}) within 50\%.

\end{enumerate}

Additional spatially  resolved observations  of ammonia would  help to
constrain  the   radial  extent  of  ammonia   (and  perhaps  vertical
structure)  and refine  our current  limits.  We  can  observe ammonia
isotopes with ALMA in Band 7 (ortho-NH$_2$D $1_{01}-0_{00}$ at 332.781
GHz  and  para-NH$_2$D $1_{01}-0_{00}$  at  332.822  GHz)  and Band  8
(ortho-NH$_2$D  $1_{10}-0_{00}$   at  470.271  GHz   and  para-NH$_2$D
$1_{10}-0_{00}$ at  494.454 GHz). We  used our models to  predict line
fluxes of about 1 Jy in band 8  and 30 mJy in band 7 using LIME in all
of our  models, and considering a  value of 0.1  for ammonia deuterium
fractionation    as   found    towards   protostellar    dense   cores
\citep{Roueff2005,Busquet2010}  and  a OPR  ammonia  of  1.  ALMA  can
detect such  line fluxes in  a couple of  hours \footnote{Calculations
  performed      with     the     ALMA      sensitivity     calculator
  (https://almascience.eso.org/proposing/sensitivity-calculator)}OA.
Observations with  JVLA or GBT of  para-ammonia (para-NH$_3$ $1_1-1_0$
at 23.694 GHz or para-NH$_3$ $2_1-2_0$ at 23.722 GHz) are not possible
since our predicted line fluxes are too small ($\sim$ 10 mJy).


\begin{acknowledgements} \emph{Herschel} is a European Space Agency space
observatory   with  science   instruments  provided   by  European-led
principal investigator consortia and with important participation from
NASA. HIFI has  been designed and built by  a consortium of institutes
and university departments from  across Europe, Canada, and the United
States under  the leadership of  SRON Netherlands Institute  for Space
Research,  Groningen, The  Netherlands, and  with  major contributions
from  Germany, France, and  the US.   Consortium members  are: Canada:
CSA,  U.  Waterloo; France:  IRAP (formerly  CESR), LAB,  LERMA, IRAM;
Germany:  KOSMA,  MPIfR,  MPS;  Ireland,  NUI  Maynooth;  Italy:  ASI,
IFSI-INAF,  Osservatorio  Astrofisico  di  Arcetri-INAF;  Netherlands:
SRON, TUD; Poland: CAMK, CBK; Spain: Observatorio Astronómico Nacional
(IGN),  Centro  de   Astrobiolog\'ia  (CSIC-INTA).   Sweden:  Chalmers
University of Technology – MC2, RSS \& GARD; Onsala Space Observatory;
Swedish  National  Space   Board,  Stockholm  University  –  Stockholm
Observatory; Switzerland:  ETH Zurich,  FHNW; USA: Caltech,  JPL, NHS.
Support  for  this  work  was  provided by  NASA  (\emph{Herschel}  OT
funding)  through  an award  issued  by  JPL/Caltech.   This work  was
partially supported  by grants  from the Netherlands  Organization for
Scientific  Research (NWO)  and  the Netherlands  Research School  for
Astronomy  (NOVA).   The  data  presented  here are  archived  at  the
\emph{Herschel}  Science Archive, http://archives.esac.esa.int/hda/ui,
under OBSID 1342198337 and 1342201585.
\end{acknowledgements}

%
%


\bibliographystyle{aa} \bibliography{bibTWHyawater}

\end{document}